\documentclass[11pt]{article}

\usepackage[margin=1in]{geometry}
\usepackage{amsmath,amssymb,amsthm,mathtools}
\usepackage{graphicx}
\usepackage{bm}
\usepackage{enumitem}
\usepackage{array}
\usepackage{natbib}
\usepackage{hyperref}
\usepackage{xcolor}
\usepackage{tikz}
\hypersetup{colorlinks=true,linkcolor=blue,citecolor=blue,urlcolor=blue}

\theoremstyle{plain}
\newtheorem{proposition}{Proposition}

\newtheorem{corollary}{Corollary}

\theoremstyle{definition}
\newtheorem{assumption}{Assumption}

\theoremstyle{remark}
\newtheorem{remark}{Remark}

\newcommand{\E}{\mathbb{E}}

\newcommand{\R}{\mathbb{R}}

\newcommand{\ATT}{\mathrm{ATT}}

\newcommand{\Cov}{\mathrm{Cov}}
\newcommand{\Var}{\mathrm{Var}}

\title{\textbf{When Covariates Mask Pre-Trends}}
\author{Parush Arora\thanks{Ashoka University. We are working on an R package that
implements the methods of this paper, which will be released soon.}}
\date{}

\providecommand{\SNplaceboOLSEarly}{}\renewcommand{\SNplaceboOLSEarly}{$0.20$}
\providecommand{\SNplaceboFlexEarly}{}\renewcommand{\SNplaceboFlexEarly}{$0.12$}
\providecommand{\SNplaceboFlexLate}{}\renewcommand{\SNplaceboFlexLate}{$-0.02$}
\providecommand{\SNlargestFlexEarly}{}\renewcommand{\SNlargestFlexEarly}{$0.36$}
\providecommand{\SNlargestFlexLate}{}\renewcommand{\SNlargestFlexLate}{$0.19$}
\providecommand{\SNtimingWald}{}\renewcommand{\SNtimingWald}{$102$}

\providecommand{\SNtabSteps}{}\renewcommand{\SNtabSteps}{10}

\begin{document}
\maketitle

\begin{abstract}
Honest difference-in-differences methods judge how far parallel trends may fail after treatment by how far it failed
before treatment, as shown by placebo estimates. When the design adjusts for covariates, the placebo estimates are
adjusted too. We show that this benchmark is valid only if the covariates are measured before the earliest placebo
window. A covariate measured later already reflects shocks to the outcome during the placebo period, so adjusting for it
hides part of the pre-trend. The placebos then look flatter than the true trend and the resulting confidence sets are
too narrow. When the covariate is also observed at an early date, the distortion can be estimated and tested. Of 90
recent published studies that adjust for time-varying covariates, 77 measure at least one covariate too late.
\end{abstract}

\clearpage

\section{Introduction}
\label{sec:intro}

A difference-in-differences (DiD) design compares how an outcome changes in treated and untreated units. It identifies
the effect of the treatment only if the two groups would have followed parallel trends without it, and that cannot be
checked after treatment, when the trend without treatment is no longer observed for the treated. The usual evidence
comes from earlier periods. Researchers run the same comparison over windows in which no unit was yet treated. If the
trends were parallel, these placebo estimates, often called pre-trends, should be close to zero.

Honest DiD methods use the placebo estimates for inference rather than as a pass-or-fail check. \citet{RR} assume that
the violation of parallel trends after treatment is at most $M$ times the largest violation seen before it, and they
report a confidence set for the effect that widens as $M$ grows. The breakdown value is the largest $M$ at which the set
still excludes zero. A breakdown value of $2$ says that the conclusion survives a violation after treatment twice as
large as the worst one before it. This relative-magnitude benchmark is now recommended practice \citep{Baker2025}.

Parallel trends is often more plausible among units that look alike, so many designs assume it only within groups of
units with the same covariates \citep{SantAnnaZhao2020,CallawaySantAnna2021}. The benchmark then compares
covariate-adjusted placebo estimates with the covariate-adjusted trend after treatment. That comparison is sound only if
holding the covariates fixed has the same effect on the trend before and after treatment. This paper shows that whether
it does depends on when the covariates are measured, and that the dating used in most applied work breaks it.

The problem arises when a covariate is measured after the start of a placebo window, for example at the end of the
pre-period or in the same year as the outcome. Such a covariate partly records what happened to the outcome during the
pre-period. The 1990 manufacturing share of a local labor market reflects how much manufacturing it lost in the 1980s.
The population of a city in a census year reflects how fast the city grew over the previous decade. Comparing treated
and untreated units with the same value of such a covariate therefore compares units whose outcomes moved alike before
treatment, and they moved alike in part because that is how they were selected. The placebo estimates are pulled toward
zero. The trend after treatment is not pulled in the same way, because the covariate was measured before the changes
after treatment took place and has not recorded them. The pre-trends then understate the violation they are meant to
bound, and the honest confidence set is too narrow.

A study from our survey of published work shows how large the effect can be. \citet{FeigenbaumGross2024} estimate how
the automation of local telephone exchanges in U.S.\ cities between 1917 and 1940 affected the employment of young women
as telephone operators. Operator employment was already rising, relative to other cities, in the cities that were about
to automate. Without covariates the two placebo coefficients of their event study are $-0.31$ and $-0.21$. The published
specification controls for city population and the number of employed young women, both measured in the census year of
the outcome. A growing city hires more operators and also gains population, so the population control absorbs the
growth that the placebo coefficients are meant to detect, and they shrink to $-0.13$ and $-0.06$. With the same
covariates measured in 1910, before every placebo window, they stay at $-0.31$ and $-0.20$. The breakdown value is
$2.15$ with the census-year covariates and $0.70$ with the 1910 covariates, so the late dating makes the result look
three times as robust. The conclusion that automation reduced operator employment holds under either dating, but its
robustness is overstated.

Section~\ref{sec:timing} makes the argument precise. Each covariate-adjusted placebo estimate splits into a trend term
and a shock term (Proposition~\ref{prop:lategen}). The trend term is the within-group trend that the benchmark is meant
to measure. The shock term is the part of the placebo estimate that comes from the shocks to the outcome that the late
covariate has recorded. The trend after treatment has no matching shock term, so the shock terms open a gap between the
pre-trends and the trend they are supposed to bound. A covariate measured before every placebo window records no shocks,
all shock terms are zero, and the benchmark is valid. The direction of the gap is predictable. In the usual case the
treatment rises with the covariate and the covariate rises with the past outcome, and the shock terms then pull every
placebo estimate down relative to the trend after treatment (Proposition~\ref{prop:latesign}). A positive trend is
partly hidden, the pre-trend check looks better than it should, and the honest set is too narrow. A negative trend gives
a conservative set instead. The direction result allows the trend to differ across windows and the shocks to be
correlated over time. In a linear design the shock terms can cancel the trend in every placebo estimate, and then no
finite $M$ gives a set that covers the effect (Proposition~\ref{prop:late}).

This problem is separate from the bias that a late covariate can cause in the estimate of the effect, which is the
concern of recent work on time-varying covariates in DiD \citep{CCPR2022,CCPS2026,BanChenKedagni2026}. That work asks when
the adjusted comparison recovers the effect, and it gives conditions under which the value of a covariate before
treatment does. We ask a different question, whether the placebo estimates remain a valid benchmark, and a covariate
measured before treatment can fail it. The two problems can occur separately. In a simple case the late covariate leaves
the estimate exactly where an early covariate would put it while shifting every placebo estimate
(Corollary~\ref{cor:contrast}). A researcher who checks only whether the estimate changes with the dating of the
covariates would find nothing wrong.

Staggered adoption makes the problem more common. Practitioners often fix time-varying covariates at their values in
the period just before treatment, which is the end of each cohort's pre-period. Section~\ref{sec:stagger} shows that the
distortion then holds cohort by cohort and survives aggregation into an event study. Regression event studies more often
measure the controls in the same period as the outcome, as in the telephone study, and each placebo coefficient then
conditions on a covariate that records the very change it measures. The \texttt{did} package of
\citet{CallawaySantAnna2021}, given time-varying covariates, uses in each comparison the value at the earlier of its two
periods. That choice dates the covariate before each window, and it removes the shock terms when the shocks are
independent over time.

The simplest remedy is to measure every covariate before the earliest placebo window. That is not always possible, and
it is not always wise, since an old measurement can describe a unit poorly. We therefore give three tools. When a
covariate is observed at an early and a late date, a test asks whether the late value predicts the changes in the
outcome during the pre-period once the treatment and the early value are taken into account
(Proposition~\ref{prop:timetest}). If it does not, the late covariate gives a valid benchmark. A rejection says that the
late covariate records the past outcome but not by how much this moves the benchmark, and a very precise covariate can
be rejected even when the effect is negligible. The second tool answers that question. The shock terms themselves are
identified from the early covariate (Proposition~\ref{prop:shock}), and they are measured in the units of the placebo
estimates, so they show directly how far the late dating moves each pre-trend. When no early measurement exists, the
third tool bounds each shock term by an identified scale times the correlation between the treatment and the window's
shock among units that share the late covariate. That correlation serves as a sensitivity parameter
(Proposition~\ref{prop:sens}). In simulations the test has correct size and detects weak dependence of the covariate on
the past outcome with binary, continuous and staggered treatments.

Late dating is the norm in applied work. We read 218 of the 270 articles published since 2022 in the five
general-interest journals and the four American Economic Journals that cite recent methods for difference-in-differences,
and 197 of them estimate such a design. Ninety adjust their main specification for time-varying covariates, and 77 of
these date at least one covariate at the end of the pre-period or in the same year as the outcome. In 71 of the 77 the
data also record the covariate at an earlier date, so the test could be run. The survey measures how common the practice
is, not how often it distorts the benchmark. That depends on how much the late covariate records the past outcome, which
is what the test and the shock terms measure in each application.

Besides the telephone study, two applications show what the tools do. The China-shock design of \citet{ADH}, with a
continuous dose, tests the theory where the data allow it most fully, because the same covariates can be measured in
1975, 1980 and 1990. All six adjusted placebo slopes are lower with the 1990 covariates than with the 1975 ones. The most
recent falls from \SNplaceboFlexEarly{} to \SNplaceboFlexLate, so a researcher with the 1990 covariates who looked at
that window would see no pre-trend at all. The estimated shock terms of the 1990 covariates are negative in every window
and account for the fall, and a sensitivity parameter of $0.1$ recovers the 1975 benchmark from the 1990 data alone. The
Medicaid-expansion example of \citet{Baker2025}, a binary treatment with covariates dated in the last pre-treatment
year, shows why the test should be read together with the shock terms. Two of its six covariates fail the test, but
their shock terms are small next to the largest placebo coefficient and its standard error, so the benchmark barely
moves. The covariate whose dating matters is the unemployment rate, which passes the test. Its early value comes from
the peak of the recession, and using it moves the estimate by more than the effect itself. The lesson is to date early
the covariates whose shock terms are large, not every covariate, and to report the estimate and the breakdown value
under each dating.

Conditioning on outcomes from before treatment is known to be delicate in DiD. Matching on lagged outcomes can induce
regression to the mean \citep{ChabeFerret2015,DawHatfield2018}, and DiD and adjustment for lagged outcomes bracket the
effect under different assumptions \citep{DingLi2019}. \citet{ZeldowHatfield2021} define confounding in DiD through
covariates whose distribution or effect changes over time, and \citet{RodriguesHatfield2026} give graphical criteria for
choosing adjustment sets, including covariates affected by earlier outcomes. Covariates that evolve with the outcome or
that the treatment affects can act as bad controls \citep{CCPR2022,CCPS2026}. \citet{BanChenKedagni2026} show that
conditioning on a mediator path with lagged outcomes changes the estimand of the DiD comparison, and \citet{CCPS2026}
give conditions under which conditioning on the value of such a covariate before treatment recovers the average
treatment effect on the treated. \citet{GSW2022,GSW2026} show that pre-trends are informative about parallel trends only
under restrictions on selection into treatment. The timing condition is a restriction of a different kind, on the
covariates used to make the pre-trends comparable. Here the pre-test passes for a mechanical reason, which adds to the
distortions from conditioning on a passed pre-test that \citet{Roth2022} documents. The condition applies to any
procedure that benchmarks conditional trends after treatment by conditional pre-trends, including the relative-magnitude
bounds of \citet{RR} applied to covariate-adjusted event studies, the pre-trend benchmark of \citet{WangEtAl2026}, and
the continuous-treatment bounds of \citet{AroraACR2026}.

Section~\ref{sec:setup} sets up honest DiD with covariates. Section~\ref{sec:timing} states the distortion, its direction
and its size in a linear design, and Section~\ref{sec:stagger} extends it to staggered adoption.
Section~\ref{sec:test} gives the test, the shock-term diagnostic and the sensitivity bound. Section~\ref{sec:mc} reports
simulations and Section~\ref{sec:apps} the survey and the three applications. Proofs are in the appendix.

\section{Honest difference-in-differences with covariates}
\label{sec:setup}

Units $i=1,\dots,n$ are treated ($D_i=1$) or untreated ($D_i=0$) at $t=1$. Treatment does not affect outcomes before it is
assigned, so changes in the outcome in the pre-period windows $s=1,\dots,K$ are untreated changes, $\Delta Y_{is}=\Delta
Y_{is}(0)$. Write $\Delta Y_{i,\mathrm{post}}(0)$ for the untreated change over the post-period window and $\Delta
Y_{i,\mathrm{post}}$ for the observed one. A covariate $\tilde X_i$ is used for adjustment.

For a random variable $W$ the within-$\tilde X$ treatment contrast is
\[
  \beta(W;\tilde X)\coloneqq\E\bigl[\E[W\mid D=1,\tilde X]-\E[W\mid D=0,\tilde X]\bigm| D=1\bigr],
\]
the difference in the conditional mean of $W$ between treated and untreated units with the same covariate, averaged over
the covariates of the treated. Write $\beta_s(\tilde X)\coloneqq\beta(\Delta Y_s(0);\tilde X)$ for the adjusted placebo
estimand of window $s$, which is identified because pre-period changes are untreated, and
$\beta_{\mathrm{post}}(\tilde X)\coloneqq\beta(\Delta Y_{\mathrm{post}}(0);\tilde X)$ for the within-$\tilde X$
differential trend of the post-period, which is not. The adjusted post-period contrast is
$m_{\tilde X}\coloneqq\beta(\Delta Y_{\mathrm{post}};\tilde X)=\ATT+\beta_{\mathrm{post}}(\tilde X)$, where $\ATT$ is the
average effect on the treated, and conditional parallel trends in its averaged form is $\beta_{\mathrm{post}}(\tilde X)=0$.

\begin{assumption}[Relative-magnitude benchmark]\label{ass:rm}
For a known $M\ge0$,
\[
  |\beta_{\mathrm{post}}(\tilde X)|\le M\max_{s\le K}|\beta_s(\tilde X)|.
\]
\end{assumption}

Assumption~\ref{ass:rm} is the relative-magnitude class of \citet{RR} written for the covariate-adjusted trends. It gives
$\ATT\in[m_{\tilde X}\pm M\max_s|\beta_s(\tilde X)|]$, and an honest band adds sampling error to the endpoints of this
interval. The restriction compares objects that hold the same covariate fixed, and it is credible when holding
$\tilde X$ fixed removes the same kind of trend before and after treatment.

The budget $M$ is measured in units of the worst pre-period violation. At $M=0$ the band maintains conditional parallel
trends. At $M=1$ it allows the post-period violation to be as large as the largest placebo estimand, and at $M=2$ twice
as large. The breakdown value $M^*$ is the largest $M$ at which the band still excludes zero. It says how large a
violation, relative to the pre-trends, the conclusion that the effect is nonzero can survive, and a larger $M^*$ makes a
result look more robust. Because $M$ multiplies the largest placebo estimand, anything that shrinks the placebo estimands
without shrinking the post-period trend raises $M^*$ above its proper value. The timing of the covariates acts through
this channel.

Applied work often reports the coefficient on the treatment in a regression with covariates. For a change in the outcome $W$
write
\[
  \beta^\omega(W;\tilde X)\coloneqq\frac{\E[\Cov(D,W\mid\tilde X)]}{\E[\Var(D\mid\tilde X)]},
\]
the partially linear coefficient of \citet{Robinson1988}, a weighted average of within-cell contrasts with the
regression weights of \citet{Angrist1998}. The same weights apply to the placebo outcomes.

\begin{remark}[Continuous treatments]\label{rem:continuous}
With a continuous dose $D\in[0,\bar d]$ the contrast is replaced by the within-$\tilde X$ slope at dose $d$,
$\beta_d(W;\tilde X)\coloneqq\E[\partial_{\tilde d}\E[W\mid D=\tilde d,\tilde X]|_{\tilde d=d}\mid D=d]$, the post-period
contrast equals the average causal response of \citet{CGBS} plus a selection term plus
$\beta_{d,\mathrm{post}}(\tilde X)$ \citep{AroraACR2026}. The selection term is nonzero when, within a covariate cell,
the units that receive larger doses would also respond more strongly to the same dose. The regression summary
$\beta^\omega$ is defined as before. All results below use only that
$W\mapsto\beta(W;\tilde X)$ is linear, so they hold for $\beta$, for $\beta_d$ at every $d$ and for $\beta^\omega$, and we
state them for a generic contrast $\beta$. Assumption~\ref{ass:lategen} below then requires the conditional means given
$(D,Z)$ to be differentiable in the dose, with differentiation and expectation interchangeable.
\end{remark}

\section{Late covariates distort the benchmark}
\label{sec:timing}

A covariate measured at the end of the pre-period partly records the changes in the outcome during the pre-period. Holding it fixed holds
part of the pre-period shocks fixed, and it does not hold the post-period shock fixed in the same way. We first state the
distortion in general and then compute its size in a linear design.

A numerical example shows the mechanism before the general statement. Return to the commuting zones of the
introduction, and suppose for simplicity that the outcome is the manufacturing share itself, so that the share at the end
of the pre-period equals its early value plus the pre-period changes. Take two zones whose share in 1990 is the same,
$20$ percent. Zone $H$ had a share of $26$ percent in 1975 and receives a large dose, since the dose rises with the share.
Zone $L$ had $22$ percent and receives a small dose. Suppose that, given its 1975 share and its dose, each zone was
expected to lose $2$ points over the pre-period, so that the two zones share the same trend. To reach $20$ percent zone
$H$ lost $6$ points and zone $L$ lost $2$, so zone $H$'s unexpected shocks add up to $-4$ points and zone $L$'s to zero.
A comparison of the two zones within the 1990 cell attributes to the dose a pre-period decline of $4$ points that no trend
produced. The shocks produced it, and the two zones share a cell only because their shocks differed. The post-period
shocks of the two zones had not yet happened when the share was measured, so the post-period comparison contains no such
term. Holding the late covariate fixed makes the high-dose zones look worse before treatment than their trends imply,
and only before treatment.

\begin{assumption}[Late covariate]\label{ass:lategen}
Units have a predetermined covariate $X_0$ and a treatment $D$. Untreated changes in the outcome in $K$ pre-period windows and in
the post-period window satisfy $\Delta Y_s(0)=\alpha_s+\tau_s(D,X_0)+u_s$ for $s\in\{1,\dots,K,\mathrm{post}\}$, where
the shocks $(u_1,\dots,u_K,u_{\mathrm{post}})$ have mean zero and are jointly independent of $(D,X_0)$. A late covariate
is $Z=\psi(X_0,V,\zeta)$ for a measurable $\psi$, with $V=\sum_{s=1}^K\Delta Y_s(0)$ the cumulative pre-period change and
$\zeta$ independent of $(D,X_0,u_1,\dots,u_K,u_{\mathrm{post}})$.
\end{assumption}

The within-cell trend $\tau_s$ may differ across windows, and the shocks may be correlated across windows. The trends are
stationary when $\tau_s=\tau$ for every $s$. The variable $\zeta$ allows the late measurement to differ from the early one
for reasons unrelated to the outcome.

\begin{proposition}[Late covariates]\label{prop:lategen}
Under Assumption~\ref{ass:lategen},
\begin{enumerate}[label=(\roman*),leftmargin=2.2em]
  \item $\beta_s(X_0)=\beta(\tau_s(D,X_0);X_0)$ for every window $s$, pre or post,
  \item $\beta_s(Z)=T_s+\varepsilon_s$ for every window, with a trend term $T_s\coloneqq\beta(\tau_s(D,X_0);Z)$ and a
  shock term $\varepsilon_s\coloneqq\beta(u_s;Z)$,
  \item $\beta_{\mathrm{post}}(Z)-\beta_s(Z)=(T_{\mathrm{post}}-T_s)+(\varepsilon_{\mathrm{post}}-\varepsilon_s)$, and
  the first term vanishes when the trends are stationary. If $Z$ does not depend on $V$, then $\varepsilon_s=0$ for every
  window. If $u_{\mathrm{post}}$ is independent of $(u_1,\dots,u_K)$, then $\varepsilon_{\mathrm{post}}=0$.
\end{enumerate}
\end{proposition}

With the early covariate, the placebo estimands measure the within-cell trends of the pre-period, which is what the
benchmark needs. With the late covariate each placebo estimand also contains the shock term $\varepsilon_s$, the treatment
contrast of the window-$s$ shock among units that share a value of the late covariate. Holding $Z$ fixed, treated units
tend to have a different $X_0$ when treatment depends on $X_0$, and their pre-period shocks must then differ to keep $Z$
fixed. So $\varepsilon_s$ is generally nonzero for the windows that enter $V$, while a post-period shock that is unrelated
to the pre-period shocks contributes nothing. The trend terms $T_s$ also differ from the within-$X_0$ trends, but with
stationary trends they move the placebo estimands and the post-period trend together and do not distort the benchmark.
In the example the late covariate places zones $H$ and $L$ in the same cell. The trend term of their comparison is zero,
because both were expected to lose $2$ points, and the shock term is $-4$ points. The 1975 share places them in
different cells, so their shocks never enter a within-cell comparison, which is part (i).

The estimate and the benchmark respond to the late covariate through different terms. The adjusted post-period contrast
moves by $T_{\mathrm{post}}-\beta_{\mathrm{post}}(X_0)+\varepsilon_{\mathrm{post}}$, which is the bias that work on bad
controls studies, while the benchmark is distorted by $\varepsilon_{\mathrm{post}}-\varepsilon_s$. Either can be zero
while the other is not.

\begin{corollary}[Estimate intact, benchmark distorted]\label{cor:contrast}
Let Assumption~\ref{ass:lategen} hold for a binary treatment with $\tau_s(D,X_0)=h_s(D)$ for every window and
$u_{\mathrm{post}}$ independent of $(u_1,\dots,u_K)$. Then $m_Z=m_{X_0}$, so the late covariate leaves the adjusted
estimate unchanged, while $\beta_s(Z)=\beta_s(X_0)+\varepsilon_s$ for every pre-period window, for the contrast $\beta$
and for the regression summary $\beta^\omega$. If $Z=\psi(X_0,\zeta)$ instead, then $\varepsilon_s=0$ in every window, the benchmark is valid, and $m_Z$ can
differ from $m_{X_0}$.
\end{corollary}

The first case is the paper's point in its simplest form. The trends depend on the treatment but not on the covariate, so
the covariate is harmless for the estimate, and it still shifts every placebo estimate. A check that compares the
estimates with an early and a late covariate finds nothing, and the pre-trend check is distorted. The second case is a
late measurement that is merely noisier, which can move the estimate and leaves the benchmark valid.

The distortion matters for coverage through a simple threshold. Consider the band of Section~\ref{sec:setup} under
Assumption~\ref{ass:rm}, built with covariate $\tilde X$. As $n\to\infty$ with $M$ fixed its coverage tends to one if
$M>M^\circ_{\tilde X}$ and to zero if $M<M^\circ_{\tilde X}$, where
\begin{equation}\label{eq:MX}
  M^\circ_{\tilde X}\coloneqq\frac{|\beta_{\mathrm{post}}(\tilde X)|}{\max_s|\beta_s(\tilde X)|},
\end{equation}
because the post-period contrast converges to the $\ATT$ plus $\beta_{\mathrm{post}}(\tilde X)$ and the half-width to $M$
times the largest placebo estimand. With stationary trends and serially independent shocks,
Proposition~\ref{prop:lategen} gives $M^\circ_{X_0}=1$ and
$M^\circ_Z=|T+\varepsilon_{\mathrm{post}}|/\max_s|T+\varepsilon_s|$, which exceeds one whenever the shock terms pull
the placebo estimands toward zero.

The direction of the distortion does not depend on linearity. For the regression summary write $T^\omega_s$ and
$\varepsilon^\omega_s$ for the trend and shock terms of Proposition~\ref{prop:lategen} with $\beta^\omega$ in place of
$\beta$.

\begin{proposition}[Direction of the distortion]\label{prop:latesign}
Let Assumption~\ref{ass:lategen} hold with $\zeta$ absent, $\E[\Var(D\mid Z)]>0$ and finite second moments, and write
$U\coloneqq\sum_{r\le K}u_r$. Suppose that (a) $Z=\Psi(A,U)$ for a scalar index $A=a(D,X_0)$ and a function $\Psi$
strictly increasing in each argument, (b) $\E[D\mid A]$ is nondecreasing in $A$, and (c)
$\E[u_s-u_{\mathrm{post}}\mid U]$ is nondecreasing in $U$ for every $s\le K$. Then
\begin{enumerate}[label=(\roman*),leftmargin=2.2em]
  \item $\varepsilon^\omega_s\le\varepsilon^\omega_{\mathrm{post}}$ for every $s\le K$, so each adjusted placebo slope is
  lowered relative to the post-period within-cell trend by $\varepsilon^\omega_{\mathrm{post}}-\varepsilon^\omega_s\ge0$
  on top of the difference $T^\omega_{\mathrm{post}}-T^\omega_s$ of the trend terms, and with stationary trends
  $\beta^\omega(\Delta Y_s(0);Z)\le\beta^\omega(\Delta Y_{\mathrm{post}}(0);Z)$,
  \item with stationary trends, if the post-period within-cell trend $\beta^\omega_{\mathrm{post}}$ is positive and
  every placebo slope exceeds $-\beta^\omega_{\mathrm{post}}$, the late covariate masks the trend, $M^\circ_Z\ge1$, and the inequality is strict when every
  $\varepsilon^\omega_s<\varepsilon^\omega_{\mathrm{post}}$, so the band at $M=1$ is too short, while if the post-period
  within-cell trend is negative, $M^\circ_Z\le1$ and the band is conservative,
  \item with $\E[D\mid A]$ nonincreasing in (b), the inequalities in (i) are reversed and the roles of the two signs in
  (ii) are exchanged.
\end{enumerate}
\end{proposition}

The conditions describe the usual case. A late measurement of a covariate rises with the covariate's early value and
with the changes in the outcome it records, which is (a), and the treatment rises with the covariate, which is (b). Condition
(c) asks each pre-period shock to move with the cumulative pre-period shock at least as much as the post-period shock
does. It holds when the post-period shock is independent of the pre-period shocks and $\E[u_s\mid U]$ is nondecreasing,
for instance with exchangeable shocks or with jointly Gaussian shocks with nonnegative covariances. It also holds for
transitory shocks in levels, whose post-period change covaries negatively with the cumulative pre-period change, and for
jointly Gaussian shocks it is equivalent to $\Cov(u_s,U)\ge\Cov(u_{\mathrm{post}},U)$, which allows persistent shocks as
long as the persistence decays. Under the conditions a late covariate pulls every placebo slope down relative to the
post-period trend. Holding the late covariate fixed, units with a larger treatment must have had smaller pre-period
shocks, and the post-period shock, which the covariate does not record, is pulled less. When the within-cell trend is
positive the pre-trends then look smaller than the trend they are meant to bound. The direction is stated relative to
the sign of the dependence of the treatment on the index $A$, so ``positive'' means in the direction of that dependence.
The linear design of Assumption~\ref{ass:linear} with jointly Gaussian variables satisfies (a)--(c) when $\chi>0$,
$\Cov(D,Z)\ge0$ and $\Xi_s\ge\Xi_{\mathrm{post}}$, and Proposition~\ref{prop:late} then gives the size of the effect.

The index $A$ depends on the treatment as well as on the early covariate because the cumulative change $V$ contains the
within-cell trends $\tau_s(D,X_0)$ along with the shocks, so a covariate that records $V$ records the treatment through
the trends. In the example $A$ is the 1975 share plus the expected change, $24$ points for zone $H$ and $20$ for zone $L$,
and $U$ is the sum of the shocks. Given $Z=A+U$, the zone with the larger index, which also has the larger dose, must
have the smaller sum of shocks, and this is the argument of the proof. Condition (a) asks the late covariate to combine
the predictable part of the pre-period changes and the shocks through one scalar index. With a single covariate this is
a mild requirement. With several covariates it asks them to record the pre-period changes through a common index, for
instance when each is a function of one score. When it fails the sign of each shock term is an empirical question, and
the estimated shock terms of Section~\ref{sec:test} answer it directly.

\subsection{A linear design}
\label{sec:linear}

\begin{assumption}[Linear design]\label{ass:linear}
Units have a predetermined covariate $X_0$ with $\Var(X_0)>0$ and a treatment $D$ that is not a linear function of $X_0$.
Untreated changes in the outcome in $K$ pre-period windows and in the post-period window are
$\Delta Y_s(0)=\alpha_s+b\,X_0+\delta\,D+u_s$ for $s\in\{1,\dots,K,\mathrm{post}\}$, where the shocks $u_s$ are uncorrelated with
$(X_0,D)$ and with each other and have mean zero and common variance $\sigma_u^2$. A late covariate is $Z=X_0+\chi V$, with
$V=\sum_{s=1}^K\Delta Y_s(0)$, $\chi\in\R$ and $\kappa\coloneqq1+\chi Kb\ne0$.
\end{assumption}

The constant $\kappa$ is the total effect of the early covariate on the late one. A unit increase in $X_0$ raises $Z$
by one directly and raises each pre-period change by $b$, hence $V$ by $Kb$ and $Z$ by $\chi Kb$.

For $\tilde X\in\{X_0,Z\}$ let $\beta_s(\tilde X)$ be the coefficient on $D$ in the linear projection of $\Delta Y_s(0)$ on
$(1,D,\tilde X)$. Under joint normality it equals $\beta^\omega(\Delta Y_s(0);\tilde X)$, and with a continuous dose it
equals the slope $\beta_d$ at every $d$.

\begin{proposition}[Linear design]\label{prop:late}
Under Assumption~\ref{ass:linear}, let $\gamma\coloneqq\Cov(D,Z)/\Var(Z)$ be the slope of the treatment on the late covariate
and $\Lambda\coloneqq\Var(D)-\Cov(D,Z)^2/\Var(Z)$ the residual variance of the treatment. Then
$\beta_s(X_0)=\delta $ in every window, and
\begin{equation}\label{eq:lateiid}
\begin{gathered}
  \beta_s(Z)=\frac{\delta }{\kappa}-\frac{\chi\gamma\sigma_u^2}{\kappa\Lambda}\ \ (s\le K),\qquad
  \beta_{\mathrm{post}}(Z)=\frac{\delta }{\kappa}+\frac{b\chi^2K\gamma\sigma_u^2}{\kappa\Lambda},\\
  \beta_{\mathrm{post}}(Z)-\beta_s(Z)=\frac{\chi\gamma\sigma_u^2}{\Lambda}.
\end{gathered}
\end{equation}
Hence $M^\circ_Z=|\delta +b\chi^2K\gamma\sigma_u^2/\Lambda|/|\delta -\chi\gamma\sigma_u^2/\Lambda|$, which diverges as $\delta $ approaches $\chi\gamma\sigma_u^2/\Lambda$.
With $b=0$, $\beta_{\mathrm{post}}(Z)=\delta =\beta_{\mathrm{post}}(X_0)$, the linear case of Corollary~\ref{cor:contrast}.
\end{proposition}

The appendix proves the proposition for an arbitrary covariance of the shocks across windows, where the shock term is
$\varepsilon_s=-\chi\gamma\,\Xi_s/\Lambda$ with $\Xi_s=\Cov(u_s,\sum_{r\le K}u_r)$, so the wedge is
$\chi\gamma(\Xi_s-\Xi_{\mathrm{post}})/\Lambda$. The late covariate absorbs part of each placebo change in proportion to
how strongly that window's shock covaries with the pre-period shocks, and part of the post-period change in proportion to
$\Xi_{\mathrm{post}}$. The wedge grows with the weight $\chi$ of the pre-period changes in the covariate and with the
dependence $\gamma$ of the treatment on it, and it vanishes at $\chi=0$. In the simulation design of Section~\ref{sec:mc}
the predetermined covariate gives adjusted placebo slopes equal to the within-cell trend $\delta =0.3$. The late covariate
lowers the placebo slope to $0.04$ while the post-period within-cell trend is $0.26$, so the band needs $M$ above six to
cover. The required $M$ can be arbitrarily large because the shock term can cancel the trend term in the placebo
estimands. When $\delta$ equals $\chi\gamma\sigma_u^2/\Lambda$ every adjusted placebo estimand is zero while the
post-period trend is not, so a band that scales the largest placebo estimand cannot cover at any finite $M$.

\section{Staggered adoption}
\label{sec:stagger}

With staggered adoption units are first treated in period $G_i\in\{2,\dots,\bar t\}$ or never, $G_i=\infty$. The building
blocks of \citet{CallawaySantAnna2021} compare cohort $g$ with the never-treated units, and with a universal base period
the placebo estimates of cohort $g$ are the adjusted contrasts of the changes before $g-1$ and the post-period estimates
those of the changes after it. Each comparison is the two-group design of Section~\ref{sec:setup} on the subpopulation
$\{G\in\{g,\infty\}\}$ with $D=\mathbf 1\{G=g\}$. Covariates that vary over time are dated in one of four ways.

\begin{enumerate}[label=(P\arabic*),leftmargin=2.6em]
  \item A predetermined value $X_1$, measured before every pre-period window of every cohort.
  \item The value in each cohort's base period, $X_{g-1}$ for every comparison of cohort $g$. This is the common practice
  of fixing covariates in the period before treatment, as in the covariate specification of \citet{Baker2025}.
  \item The value at the start of each comparison's window. The \texttt{did} package of \citet{CallawaySantAnna2021},
  given time-varying covariates, takes in each comparison the value in the earlier of its two periods (version 2.1.2,
  through \texttt{panel2cs2}), which is this dating.
  \item The value in the period of each observation, $X_t$ for the outcome in period $t$. This is the usual practice in
  regression event studies with time-varying controls.
\end{enumerate}

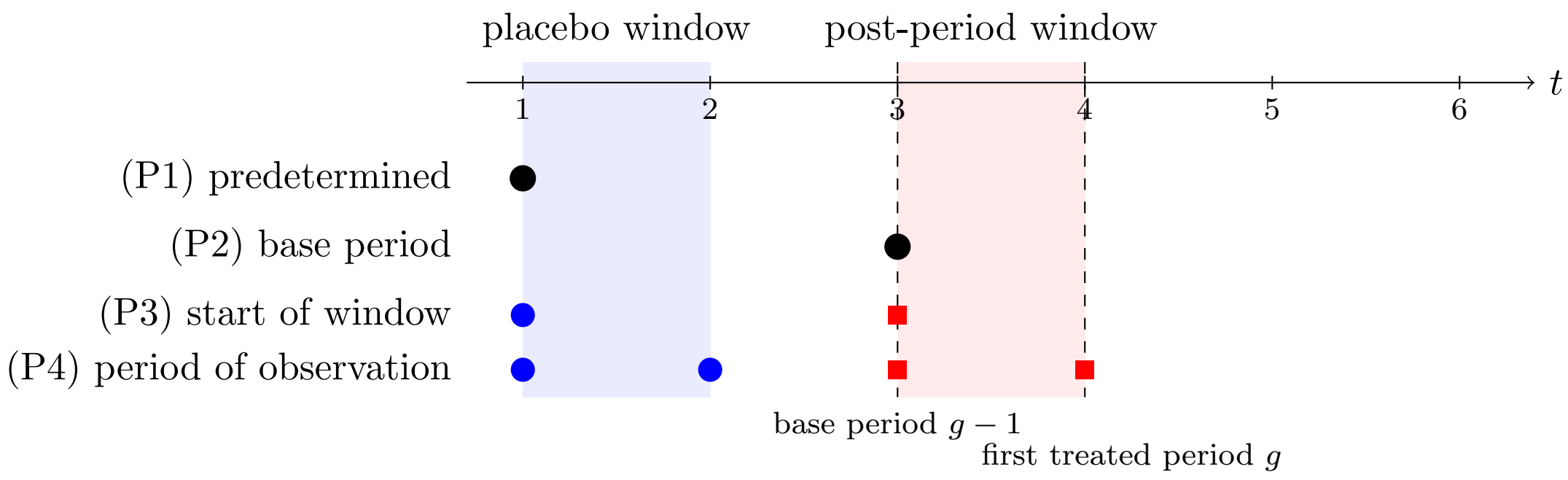
\begin{figure}[t]
\centering
\begin{tikzpicture}[x=1.7cm,y=0.62cm,font=\small]
  \fill[blue!8] (1,0.3) rectangle (2,-4.6);
  \fill[red!8] (3,0.3) rectangle (4,-4.6);
  \node[above] at (1.5,0.3) {placebo window};
  \node[above] at (3.5,0.3) {post-period window};
  \draw[->] (0.7,0) -- (6.4,0) node[right] {$t$};
  \foreach \t in {1,...,6} {\draw (\t,0.1) -- (\t,-0.1); \node[above] at (\t,-0.75) {\scriptsize$\t$};}
  \draw[dashed] (3,0.3) -- (3,-4.6); \draw[dashed] (4,0.3) -- (4,-4.6);
  \node[below] at (3,-4.6) {\scriptsize base period $g-1$};
  \node[below] at (4.25,-5.05) {\scriptsize first treated period $g$};
  \node[left] at (0.7,-1.4) {(P1) predetermined};
  \node[left] at (0.7,-2.4) {(P2) base period};
  \node[left] at (0.7,-3.4) {(P3) start of window};
  \node[left] at (0.7,-4.2) {(P4) period of observation};
  \tikzset{both/.style={circle,fill=black,inner sep=2.4pt},pre/.style={circle,fill=blue,inner sep=2.2pt},
    post/.style={rectangle,fill=red,inner sep=2.4pt}}
  \node[both] at (1,-1.4) {};
  \node[both] at (3,-2.4) {};
  \node[pre] at (1,-3.4) {}; \node[post] at (3,-3.4) {};
  \node[pre] at (1,-4.2) {}; \node[pre] at (2,-4.2) {};
  \node[post] at (3,-4.2) {}; \node[post] at (4,-4.2) {};
\end{tikzpicture}
\caption{The four datings of a time-varying covariate for cohort $g=4$, shown for the placebo window from period $1$ to
period $2$ and for the post-period window from period $3$ to period $4$. Black dots mark a value used for both windows,
blue dots values used for the placebo window and red squares values used for the post-period window. A value measured
after the start of the placebo window records part of the change in the outcome over that window.}
\label{fig:dating}
\end{figure}

Figure~\ref{fig:dating} shows the four datings for one cohort. Let the covariate evolve as $X_t=\psi_t(X_1,Y_t(0)-Y_1(0),\zeta_t)$, so that it records the cumulative untreated change up
to period $t$, with $\zeta_t$ independent of everything else, and let Assumption~\ref{ass:lategen} hold within each
comparison, with the one-period changes as windows and $X_1$ in the role of $X_0$. Event-study aggregates average the
cohort contrasts at each event time with nonnegative weights.

\begin{proposition}[Staggered adoption]\label{prop:stagger}
\begin{enumerate}[label=(\roman*),leftmargin=2.2em]
  \item Under (P1) the benchmark is valid for every cohort, as in Proposition~\ref{prop:lategen}(i).
  \item Under (P2) Propositions~\ref{prop:lategen} and~\ref{prop:latesign} hold cohort by cohort with $Z=X_{g-1}$ and
  $V$ the cumulative change of the cohort's pre-period. If the conditions of Proposition~\ref{prop:latesign} hold in
  every cohort with stationary trends, every event-study placebo aggregate lies weakly below the post-period aggregate
  trend formed with the same cohort weights.
  \item Under (P3), if the shocks are independent across periods, every shock term is zero, because the covariate at the
  start of a window records only earlier shocks. The trend terms are $T_t=\beta(\tau_t(D,X_1);X_{t-1})$ for the window
  ending at $t$, and they differ across windows because the conditioning covariate does, so the benchmark compares
  trends within different cells. With serially correlated shocks the shock terms are nonzero again.
\end{enumerate}
\end{proposition}

Dating (P2) gives each cohort a different late covariate, and a later cohort's covariate records a longer pre-period.
Dating (P3) avoids the shock terms with serially independent shocks but changes the conditioning variable from window to
window, and the band is then valid under a relative-magnitude restriction on trends measured within different cells. In
the simulations of Section~\ref{sec:mc} the change of cells is minor, and dating (P3) covers while dating (P2) does not.
Dating (P4) is not covered by the proposition. The covariate changes with the observation rather than with the
comparison, and the regression event studies that use it are not built from the contrasts of Section~\ref{sec:setup}.
The logic of Proposition~\ref{prop:lategen} still applies window by window. A placebo coefficient compares the outcome
in a pre-treatment period with the outcome in the omitted period, and with unit fixed effects the regression holds fixed
the change in the covariate between those two periods. That change records the change in the outcome that the placebo
coefficient measures, as the late covariate of Assumption~\ref{ass:lategen} does, so each placebo coefficient carries a
shock term. The regression averages such comparisons across cohorts and control units with weights of its own, which
can be negative with staggered timing, so the theory does not sign the total shift. The shock terms of
Section~\ref{sec:test}, estimated with the same regression, measure it directly, and Section~\ref{sec:phone} does this
for a published study. A predetermined value avoids all of these issues when it exists.

\section{Testing, measuring and bounding the distortion}
\label{sec:test}

Comparing adjusted placebo slopes computed with an early and a late version of a covariate does not by itself reveal the
distortion. A later measurement that is noisier, or that differs from the early one for reasons unrelated to the outcome,
moves the trend terms of Proposition~\ref{prop:lategen}, which shift every window alike and leave the benchmark valid. The
distortion comes from the shock terms, and these are present only if the late covariate predicts the pre-period changes
beyond what the treatment and the early covariate predict.

\begin{proposition}[A test of covariate timing]\label{prop:timetest}
Let Assumption~\ref{ass:lategen} hold and suppose $\E[\Var(Z\mid D,X_0)]>0$. Consider the null hypothesis
\[
  H_0:\quad \E[\Delta Y_s\mid D,X_0,Z]=\E[\Delta Y_s\mid D,X_0]\quad\text{for every pre-period window }s\le K.
\]
\begin{enumerate}[label=(\roman*),leftmargin=2.2em]
  \item If $Z$ does not depend on $V$, then $H_0$ holds.
  \item Under $H_0$, $\varepsilon_s=0$ for every $s\le K$, so every adjusted placebo estimand with $Z$ equals its trend
  term. If in addition $u_{\mathrm{post}}$ is independent of $(u_1,\dots,u_K)$ and the trends are stationary, then
  $M^\circ_Z=1$, so $Z$ gives the same valid benchmark as $X_0$.
  \item Under $H_0$ the partially linear coefficient $\vartheta_s\coloneqq\E[(Z-\E[Z\mid D,X_0])\Delta Y_s]/\E[\Var(Z\mid
  D,X_0)]$ is zero for every $s\le K$. In the linear design of Assumption~\ref{ass:linear} with $Z=X_0+\chi V+\zeta$,
  \[
    \vartheta_s=\frac{\chi\,\Xi_s}{\chi^2\Var(U)+\Var(\zeta)},
  \]
  which is nonzero whenever $Z$ records the pre-period changes ($\chi\ne0$) and the shock of window $s$ covaries with
  them ($\Xi_s\ne0$).
  \item A nonzero shock term in any pre-period window implies that $H_0$ fails. The converse is false. In the linear design
  with $\Cov(D,Z)=0$ every shock term is zero while $\vartheta_s\ne0$.
\end{enumerate}
\end{proposition}

The coefficients $\vartheta_s$ are estimated by a cross-fitted partially linear regression \citep{Robinson1988,DML2018}
with $Z$ in the role of the treatment and $(D,X_0)$ as controls, and their influence functions give a joint Wald test over
windows. By part (iv) every alternative that distorts the benchmark violates $H_0$. The coefficients $\vartheta_s$
detect the violations that are correlated with the residual of $Z$, and in the linear design these are all of them,
since a nonzero shock term $-\chi\gamma\Xi_s/\Lambda$ requires $\chi\Xi_s\ne0$ and hence $\vartheta_s\ne0$. Outside the
linear design $\vartheta_s$ can be zero while $H_0$ fails, and adding functions of $Z$, such as its square, to the test
raises its power against such alternatives. The test can also reject when the distortion is zero. In the linear design its $t$-statistic grows with $\sqrt n$ times
$\chi\,\Xi_s/(\sigma_u\{\chi^2\Var(U)+\Var(\zeta)\}^{1/2})$, so a precisely measured late covariate is detected at small
$\chi$, where the distortion is also small. A rejection is a reason to look at the size of the shock terms, and the early
covariate identifies them.

\begin{proposition}[The shock terms are identified]\label{prop:shock}
Under Assumption~\ref{ass:lategen}, let $R_s\coloneqq\Delta Y_s-\E[\Delta Y_s\mid D,X_0]$ for $s\le K$. Then $R_s=u_s$,
and the shock term of window $s$ is the adjusted placebo estimand of the residual change, $\varepsilon_s=\beta(R_s;Z)$.
The trend term $T_s=\beta_s(Z)-\varepsilon_s$ is identified as well.
\end{proposition}

The estimate of $\varepsilon_s$ is the cross-fitted partially linear slope of the residual change on the treatment given
$Z$, with the residual itself cross-fitted on $(D,X_0)$. It is in the units of the placebo slopes, so it says directly how
far the late covariate moves each benchmark. The post-period shock term is not identified, because the post-period change
contains the treatment effect, and condition (c) of Proposition~\ref{prop:latesign} or the independence in
Proposition~\ref{prop:lategen}(iii) is what signs or removes it.

When no early measurement exists, neither the test nor the diagnostic can be computed. The shock term can still be
bounded. Write $\tilde D\coloneqq D-\E[D\mid Z]$.

\begin{proposition}[A sensitivity bound]\label{prop:sens}
Under Assumption~\ref{ass:lategen} with $\E[\Var(D\mid Z)]>0$ and finite second moments, for every $s\le K$,
\[
  \varepsilon^\omega_s=\rho_s\,\frac{\mathrm{sd}(u_s-\E[u_s\mid Z])}{\mathrm{sd}(\tilde D)},\qquad
  |\varepsilon^\omega_s|\le|\rho_s|\,B_s,\qquad
  B_s\coloneqq\frac{\mathrm{sd}(\Delta Y_s-\E[\Delta Y_s\mid D])}{\mathrm{sd}(\tilde D)},
\]
where $\rho_s\coloneqq\mathrm{Corr}(\tilde D,u_s-\E[u_s\mid Z])$ is the correlation between the treatment and the window-$s$
shock among units that share the late covariate. The scale $B_s$ is identified from $(\Delta Y_s,D,Z)$.
\end{proposition}

The shocks are independent of the treatment, so $\rho_s$ is nonzero only because conditioning on $Z$ links them, and it
measures how strongly the late covariate records the outcome in the direction of the treatment. Under the conditions of
Proposition~\ref{prop:latesign} with $\varepsilon^\omega_{\mathrm{post}}=0$, a bound $|\rho_s|\le\bar\rho$ places the
shock-free placebo slope $T^\omega_s$ in $[\beta^\omega_s(Z),\beta^\omega_s(Z)+\bar\rho B_s]$, and the band is formed with
the larger of $|\beta^\omega_s(Z)|$ and $|\beta^\omega_s(Z)+\bar\rho B_s|$ in place of $|\beta^\omega_s(Z)|$. Reporting
the band as a function of $\bar\rho$, and the value of $\bar\rho$ at which its conclusion changes, turns the timing
condition into a sensitivity analysis. The bound is conservative, since $B_s$ uses the variance of the pre-period change
given the treatment alone, which exceeds the shock variance. The correction moves each placebo slope in one direction
only, upward, because under the conditions of Proposition~\ref{prop:latesign} the shock term can only have lowered it.
Without those conditions the band should use $|\beta^\omega_s(Z)|+\bar\rho B_s$.

Three sources inform the choice of $\bar\rho$. The first is the bound itself read backward. Since
$|\varepsilon^\omega_s|\le|\rho_s|B_s$, a shock term of size $e$ requires $|\rho_s|\ge e/B_s$, so a researcher who can
name the size of shock term that would matter, such as the standard error of the largest placebo estimate, can translate
it into the value of $\bar\rho$ that it needs. The second is a covariate observed at two dates in the same data. Its
estimated shock terms give ratios $|\hat\varepsilon_s|/B_s$, and their largest value is a benchmark for covariates of a
similar kind that are observed only late, in the way that observed covariates calibrate unobserved ones in other
sensitivity analyses. The third is a related design in which an early date exists. In the China application of
Section~\ref{sec:china} the ratios lie between $0.04$ and $0.16$.

The results suggest a simple practice. Date every covariate before the earliest pre-period window when the data allow it.
When a covariate is available only at the end of the pre-period but also at an earlier date, test whether its late value
predicts the pre-period changes and estimate the shock terms. Date early the covariates whose shock terms are large
relative to the largest placebo estimate, and keep the late values of the others, since a stale measurement has costs of
its own. A shock term is large when it is of the order of the standard error of the largest placebo estimate or more.
A shift of that size can change which window sets the benchmark and moves the breakdown value by as much as sampling
error does, while a shock term well below that standard error moves the band by less than its own uncertainty. In the
applications the shock terms of the 1990 China covariates are about one such standard error, those of the census-year
automation covariates three to four, and those of the 2013 Medicaid covariates a tenth to a quarter of one. When no early date exists, report the sensitivity bound. In every case report the estimate, the largest
placebo estimate and the breakdown value under each dating, because the dating can move the estimate without moving the pre-trends.
Appendix~\ref{app:steps} sets out these steps in the order in which they are carried out.

\section{Monte Carlo evidence}
\label{sec:mc}

The first design is an instance of Assumption~\ref{ass:linear}. Units have $X_0\sim N(0,1)$
and treatment $D=0.8X_0+\nu$ with $\nu\sim N(0,1)$. Untreated changes in two pre-period windows and one post-period window
are $X_0+gD+u_s$ with $u_s\sim N(0,1)$. The post-period change adds a homogeneous effect $-0.5\,D$. The late covariate is
$X_{\mathrm{end}}=X_0+0.5\,(\Delta Y_{1}(0)+\Delta Y_{2}(0))$. Each estimator is an OLS slope on $(1,D,X)$ with $n=700$
and $2{,}000$ replications, and the bands use the joint covariance of the post-period and placebo slopes. With $X_0$ the
adjusted placebo slope recovers $\delta $, the band that maintains conditional parallel trends ($M=0$) fails to cover when
$\delta =0.3$, and the band with $M=1$ covers (Table~\ref{tab:mc}, first panel). With $X_{\mathrm{end}}$ and $\delta =0.3$ the
adjusted placebo slope falls to $0.04$, the value that Proposition~\ref{prop:late} gives, and the $M=1$ band covers in
$5\%$ of replications. With $\delta =0$ the proposition gives a placebo slope of $-0.08$ and a post-period within-cell trend of
$+0.08$. The $M=0$ band then covers in only $42\%$ of replications, and the $M=1$ band sits at the boundary $M^\circ_Z=1$
of \eqref{eq:MX} and covers.

The second design adds normal measurement noise $\zeta$ with mean zero and variance $0.25$ to the late covariate,
$Z=X_0+\zeta+\chi V$, so that $\chi=0$ is a later measurement that records no pre-period outcome, and it also runs with
a binary treatment $D=\mathbf 1\{0.8X_0+\nu>0\}$. The test of Proposition~\ref{prop:timetest}, a joint Wald test on the two
pre-period windows, rejects in $4.7\%$ to $5.6\%$ of replications at $\chi=0$, in at least $93\%$ at $\chi=0.05$ and in
all replications from $\chi=0.1$, with either treatment (second panel). At $\chi=0$ the noisy covariate controls for $X_0$ only in
part. The part of $X_0$ it misses is correlated with the treatment and loads on $D$, which raises the adjusted placebo
slope from $0.30$ to $0.44$. It raises the post-period trend by the same amount, so the band still covers. This is the
shift of the trend term in Proposition~\ref{prop:lategen}. At $\chi=0.1$ the test always rejects while the band built with $Z$ still covers in
$91\%$ to $95\%$ of replications, the case in which the size of the shock terms is the more useful report. At $\chi=0.5$
the band built with $Z$ covers in $9\%$ of replications with a continuous dose and in $45\%$ with a binary treatment.

The third design has periods $1,\dots,6$, a cohort treated from period 4, a cohort treated
from period 6 and never-treated units, assigned by thirds of $0.8X_1+\nu$, with $n=1{,}500$ and $1{,}000$ replications.
Untreated one-period changes are $X_1+\delta\,\mathbf 1\{G<\infty\}+u_t$ with independent standard normal shocks, the effect
is $-0.5$ in the first treated period, and the covariate is $X_t=X_1+0.3\,(Y_t(0)-Y_1(0))$. Each cohort is compared with
the never-treated units by regression adjustment, and the cohort estimates are averaged with cohort-size weights. The
placebos are the changes into event times $-2$ and $-1$ and the target is the effect at event time $0$. With $\delta =0.3$ the
base-period dating (P2) lowers the aggregate placebo slopes from $0.30$ to $0.02$ while the post-period trend stays at
$0.27$, and the $M=1$ band covers in $13\%$ of replications (third panel). The start-of-window dating (P3) gives placebo
slopes of $0.28$ and covers.

\begin{table}[t]
\centering
\small
\begin{tabular}{llccc}
\hline
\multicolumn{5}{l}{\emph{Timing design: coverage of the effect (target $95\%$)}}\\
within-cell trend & covariate & adjusted placebo & $M=0$ & $M=1$ \\
\hline
$\delta =0$ & $X_0$ & $0.00$ & $95\%$ & $100\%$ \\
$\delta =0$ & $X_{\mathrm{end}}$ & $-0.08$ & $42\%$ & $100\%$ \\
$\delta =0.3$ & $X_0$ & $0.30$ & $0\%$ & $100\%$ \\
$\delta =0.3$ & $X_{\mathrm{end}}$ & $0.04$ & $0\%$ & $5\%$ \\
\hline
\multicolumn{5}{l}{\emph{Timing test, $\delta =0.3$: rejection rate and the band built with the late covariate $Z$}}\\
treatment & weight $\chi$ & adjusted placebo & test rejects & $M=1$ coverage \\
\hline
continuous & $0$ & $0.44$ & $4.7\%$ & $99.6\%$ \\
continuous & $0.1$ & $0.32$ & $100\%$ & $91.2\%$ \\
continuous & $0.5$ & $0.10$ & $100\%$ & $8.6\%$ \\
binary & $0$ & $0.55$ & $5.0\%$ & $99.2\%$ \\
binary & $0.1$ & $0.37$ & $100\%$ & $95.3\%$ \\
binary & $0.5$ & $0.07$ & $100\%$ & $44.7\%$ \\
\hline
\multicolumn{5}{l}{\emph{Staggered design, $\delta =0.3$: aggregate placebo slopes and coverage}}\\
dating & placebos ($e=-2,-1$) & post-period trend & $M=0$ & $M=1$ \\
\hline
(P1) predetermined & $0.30$, $0.30$ & $0.30$ & $0\%$ & $99.7\%$ \\
(P2) base period & $0.02$, $0.02$ & $0.27$ & $0\%$ & $13.2\%$ \\
(P3) start of window & $0.28$, $0.28$ & $0.27$ & $0\%$ & $99.9\%$ \\
\hline
\end{tabular}
\caption{Monte Carlo results. First panel, timing design ($n=700$, $2{,}000$ replications), mean adjusted placebo slope in
the last pre-period window and coverage of the covariate band for two trend budgets. Second panel, the test of
Proposition~\ref{prop:timetest} at the $5\%$ level with $Z=X_0+\zeta+\chi V$ ($n=700$, $2{,}000$ replications), the mean
adjusted placebo slope with $Z$ in the last window, and the coverage of the $M=1$ band built with $Z$. With $X_0$ the
adjusted placebo slope is $0.30$ and the band covers in at least $99\%$ of replications. Third panel, staggered design
($n=1{,}500$, $1{,}000$ replications), mean aggregate placebo slopes, mean aggregate post-period within-cell trend and
coverage of the effect at event time $0$. With $\delta =0$ the (P2) band at $M=1$ covers in $99.9\%$ and the $M=0$ band in
$57\%$.}
\label{tab:mc}
\end{table}

\section{Applications}
\label{sec:apps}

Before turning to the applications we ask how often published work dates covariates after a placebo window. The frame is every
article published from 2022 onward in the \textit{American Economic Review}, \textit{Econometrica}, the \textit{Journal
of Political Economy}, the \textit{Quarterly Journal of Economics}, the \textit{Review of Economic Studies} and the four
\textit{American Economic Journals} that cites at least one of eight methods papers on difference-in-differences and event
studies, 270 articles in all.\footnote{The eight are \citet{CallawaySantAnna2021}, \citet{RR}, \citet{Roth2022} and the
papers of Sun and Abraham (2021), de Chaisemartin and D'Haultf\oe uille (2020), Borusyak, Jaravel and Spiess (2024),
Goodman-Bacon (2021) and Sant'Anna and Zhao (2020). The frame was drawn from OpenAlex and fixed before any article was
read. Each article was coded from its text and appendix for the dating of its time-varying covariates, whether an earlier
value is recorded in its data, and how it uses the pre-trends.} We could obtain the text of 218 of them, and the other 52 had no copy that we could access. Of the 218, 197 estimate a
difference-in-differences or event-study design. Of these, 90 adjust the main specification for time-varying covariates.
Twelve date every such covariate before the earliest pre-period window, as the condition requires. The other 77 date at
least one covariate at the end of the pre-period or in the same year as the outcome, dating (P2) or (P4), so the
covariate postdates at least one placebo window, and one more dates covariates at the start of each window. In 71 of
the 77 the data also record the covariate at an earlier date, so the test of Proposition~\ref{prop:timetest} and the
shock terms of Proposition~\ref{prop:shock} can be computed. Six of the 197 articles report honest bands, and five of
these six use late covariates in the main specification. Late dating is therefore the norm, and an early value is
usually at hand. The survey codes the dating from the text and does not reestimate the studies, so it shows that the
condition is often left unchecked, not that it fails. Whether a late covariate distorts the benchmark depends on how much
it records the pre-period changes, and the three applications show that the answer varies from case to case.

\subsection{The China shock, a continuous dose}
\label{sec:china}

The first application is the effect of Chinese import competition on manufacturing employment in U.S.\ commuting zones
(CZs) \citep{ADH}. The dose is the change in Chinese import exposure per worker, in thousands of dollars, over 1990--2000
and 2000--2007. The outcome is manufacturing employment per 100 residents aged 15--64, from the imputed County Business
Patterns (CBP) of \citet{EFSY} and SEER population counts, aggregated to CZs with the crosswalk of \citet{ADH}. The two
periods are stacked, with the 2000--2007 change multiplied by $10/7$ so that both are ten-year changes as in
\citet{ADH}, and the regressions include a period effect, population weights and standard errors clustered by state.
The placebo outcomes are the six ten-year changes from 1975--1985 to 1980--1990, each paired with the period's own dose.
The covariates are the nine SIC division shares of CBP employment, log employment and region. Because CBP is annual from
1975, the same covariate set can be measured in 1975, before every placebo window, in 1980, or in 1990, at the end of the
pre-period, which is when \citet{ADH} measure their controls for the first period. Every slope is a cross-fitted
partially linear slope with lasso first stages on splines of the covariates and their pairwise interactions, the median
over $25$ splits with folds formed by state.

\begin{table}[t]
\centering
\small
\begin{tabular}{lccc}
\hline
 & \multicolumn{3}{c}{covariates measured in} \\
 & 1975 & 1980 & 1990 \\
\hline
\multicolumn{4}{l}{\emph{Adjusted placebo slopes}}\\
1975--1985 & $0.363$ $(0.125)$ & $0.331$ $(0.147)$ & $0.191$ $(0.116)$ \\
1976--1986 & $0.261$ $(0.080)$ & $0.245$ $(0.099)$ & $0.128$ $(0.073)$ \\
1977--1987 & $0.272$ $(0.076)$ & $0.250$ $(0.075)$ & $0.149$ $(0.062)$ \\
1978--1988 & $0.274$ $(0.073)$ & $0.267$ $(0.071)$ & $0.186$ $(0.061)$ \\
1979--1989 & $0.223$ $(0.057)$ & $0.201$ $(0.056)$ & $0.102$ $(0.059)$ \\
1980--1990 & $0.117$ $(0.079)$ & $0.142$ $(0.093)$ & $-0.022$ $(0.105)$ \\
\hline
\multicolumn{4}{l}{\emph{Post-period slope and honest bands}}\\
adjusted slope $m_{\tilde X}$ & $-0.356$ $(0.049)$ & $-0.316$ $(0.045)$ & $-0.274$ $(0.045)$ \\
band at $M=0$ & $[-0.45,\,-0.26]$ & $[-0.41,\,-0.23]$ & $[-0.36,\,-0.19]$ \\
band at $M=1$ & $[-0.99,\,0.26]$ & $[-0.97,\,0.30]$ & $[-0.71,\,0.16]$ \\
breakdown value $M^*$ & $0.55$ & $0.49$ & $0.60$ \\
\hline
\multicolumn{4}{l}{\emph{Shock terms of the late covariates (Proposition~\ref{prop:shock}), early covariates from 1975}}\\
1975--1985 & & $-0.02$ $(0.18)$ & $-0.11$ $(0.18)$ \\
1976--1986 & & $-0.05$ $(0.12)$ & $-0.18$ $(0.12)$ \\
1977--1987 & & $-0.01$ $(0.09)$ & $-0.10$ $(0.11)$ \\
1978--1988 & & $0.00$ $(0.09)$ & $-0.06$ $(0.10)$ \\
1979--1989 & & $-0.03$ $(0.08)$ & $-0.12$ $(0.09)$ \\
1980--1990 & & $-0.07$ $(0.13)$ & $-0.24$ $(0.16)$ \\
\hline
\end{tabular}
\caption{The China-shock design by the date at which the same covariates are measured. Stacked 1990--2000 and 2000--2007
periods, CBP outcome, each placebo change paired with the period's dose, state-clustered standard errors in parentheses.
Cross-fitted partially linear slopes with flexible controls. With OLS and linear controls the 1980--1990 slope with
covariates from 1975 is \SNplaceboOLSEarly{} \citep[Online Table~\SNtabSteps, row~5]{AroraACR2026}. Bands use the relative-magnitude class with the largest
absolute placebo slope and the union of the vertex intervals, and $M^*$ is the largest $M$ at which the band excludes
zero. Shock-term standard errors treat the first-step residuals as known.}
\label{tab:china}
\end{table}

\begin{figure}[t]
\centering
\includegraphics[width=0.85\textwidth]{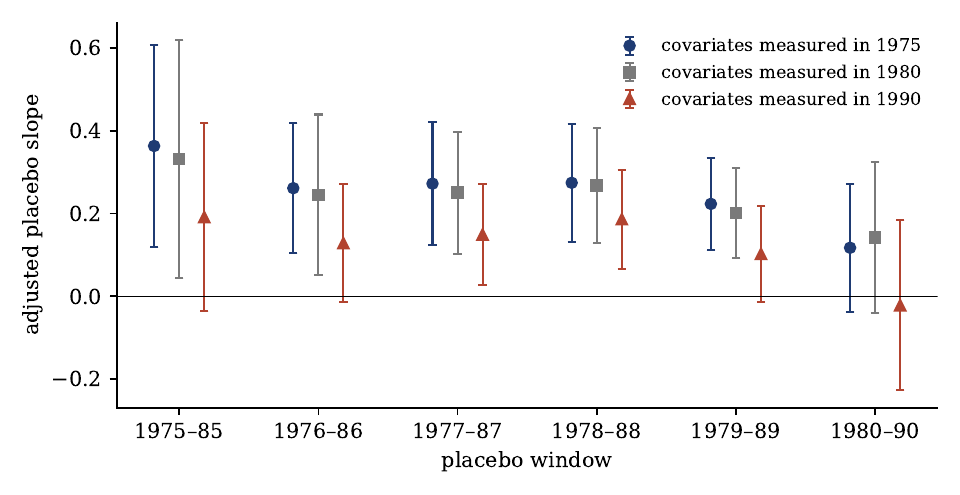}
\caption{Adjusted placebo slopes in the China-shock design with the same covariates measured in 1975, 1980 and 1990,
with $95\%$ confidence intervals. The values are those of Table~\ref{tab:china}.}
\label{fig:china}
\end{figure}

Table~\ref{tab:china} reports the adjusted placebo slopes, and Figure~\ref{fig:china} plots them. With the predetermined 1975 covariates
all six are positive, five of them by more than two standard errors. Every one is lower with the 1990 covariates than with
the 1975 covariates, by $0.09$ to $0.17$, and eleven of the twelve comparisons between adjacent dates go the same way.
The exception, the most recent window with 1980 covariates, moves by less than a third of its standard error. The
largest slope falls from \SNlargestFlexEarly{} to \SNlargestFlexLate{} and the most recent from \SNplaceboFlexEarly{} to \SNplaceboFlexLate. A researcher with the 1990
covariates who checked the most recent window would see no pre-trend at all.

The test of Proposition~\ref{prop:timetest} points to the cause. Given the dose and the 1975 covariates, the 1990 manufacturing share predicts
the change in every one of the six windows, with $t$-statistics between $6.1$ and $8.5$ and a joint Wald statistic of
\SNtimingWald{} on six degrees of freedom. The ten 1990 covariates jointly predict the 1980--1990 change, with a Wald statistic of
$149$ on ten degrees of freedom. The 1990 manufacturing share partly records the manufacturing decline of the 1980s, which
is the mechanism of Section~\ref{sec:timing}. The estimated shock terms of the 1990 covariates are negative in every
window, between $-0.06$ and $-0.24$. Their average, $-0.14$, is close to the average fall of the placebo slopes between the
1975 and 1990 columns, $-0.13$, so on average the trend terms barely move and the late dating changes the benchmark
through the shock terms. The shock terms of the 1980 covariates
are close to zero, as are the changes between the 1975 and 1980 columns. The individual shock terms are imprecise, but
their signs agree in eleven of twelve cases with Proposition~\ref{prop:latesign}, whose conditions are plausible here.
The dose rises with the manufacturing share, and the share rises with its earlier value and with the employment changes
it records. With several covariates condition (a) holds at best approximately, and the estimated shock terms, which do
not rely on it, carry the argument.

The late dating also moves the estimate. The adjusted post-period slope is $-0.36$ with 1975
covariates and $-0.27$ with 1990 covariates, so a covariate that records the 1980s decline absorbs part of the
association between the dose and the later decline. The honest band then looks more robust with the late covariates, with
a breakdown value of $0.60$ against $0.55$, because its benchmark is smaller by more than its estimate is. All three bands
at $M=1$ contain zero. The 1990 band, from $-0.71$ to $0.16$, is about a third shorter than the 1975 band.

The sensitivity bound shows what a researcher without an early date could have concluded. The ADH public files contain the covariates only at the start of each
period, so a researcher using them has the 1990 dating and no early one. The sensitivity bound of
Proposition~\ref{prop:sens} gives scales $B_s$ between $1.37$ and $1.46$. At $\bar\rho=0.1$ the largest corrected placebo slope is
$0.34$, close to the $0.36$ that the 1975 covariates give, and the breakdown value falls from $0.60$ to $0.43$. At
$\bar\rho=0.05$ it is $0.50$. The estimated shock terms correspond to values of $|\varepsilon_s|/B_s$ between $0.04$ and
$0.16$, the smallest values of $\bar\rho$ that allow shock terms of those sizes, so a researcher who allowed a correlation of about $0.1$ would have recovered the predetermined benchmark from the
late data alone.

\subsection{Medicaid expansion, a binary treatment}
\label{sec:medicaid}

The second application uses the running example of \citet{Baker2025}, the effect of the 2014 Medicaid expansions on
county adult mortality, a binary treatment whose covariates are dated in the last pre-treatment year. It shows how the shock terms separate a
rejection that moves the benchmark from one that does not, and it shows that dating every covariate early is not the
remedy.

We use the replication data of \citet{Baker2025} and follow their sample construction and weights. The
outcome is the crude mortality rate of adults aged 20--64 per 100,000, 2009--2019. The treated group is the 978 counties
in states that expanded Medicaid in 2014 and the comparison group the 1,222 counties in states that had not expanded by
2019, in 39 states. Observations are weighted by the county's adult population in 2013. The covariates are the shares of
the adult population that are female, white and Hispanic, the unemployment rate, the poverty rate and median household
income. In the covariate-adjusted event study of \citet{Baker2025}, the estimator of \citet{CallawaySantAnna2021} with a
universal base period takes these covariates in the base year 2013, the end of the pre-period 2009--2013, which is dating
(P2). The data also record them in 2009, before every pre-period window. Event-time effects are estimated by outcome
regression with the base year 2013, so the placebo coefficients are the changes $Y_t-Y_{2013}$ for
$t=2009,\dots,2012$. A placebo coefficient of this kind is minus the trend from year $t$ to the base year. A negative
coefficient is a rising pre-trend, and a shock term that lowers the trend, as Proposition~\ref{prop:latesign} predicts,
raises the coefficient. We report shock terms on the scale of the coefficients in this application and the next.\footnote{The doubly robust estimator used by \citet{Baker2025} with all adoption cohorts does not
converge in this two-group sample with population weights, since the propensity-score logit reaches fitted probabilities
of zero or one.}

We first test the 2013 covariates. Each annual pre-period change from 2009--2010 to 2012--2013 is regressed on
treatment, the 2009 covariates and their squares, and the six 2013 covariates, with population weights. The joint Wald
test that the 24 coefficients on the 2013 covariates are zero gives $48.0$ on 24 degrees of freedom ($p=0.003$) with
county-clustered standard errors. With 24 restrictions and 39 states a state-clustered version is unreliable, and we do
not report it. Table~\ref{tab:medicaid}, upper panel, splits the statistic by covariate. The 2013 female share ($p=0.006$)
and Hispanic share ($p=0.028$) predict the pre-period changes given their 2009 values. The female share of adults records
mortality mechanically, since deaths among men raise it. The 2013 unemployment rate ($p=0.83$) and poverty rate ($p=0.91$)
do not. The shock terms of Proposition~\ref{prop:shock} show that the rejection matters little for the benchmark. With all
six covariates dated in 2013 they are between $0.2$ and $0.6$ deaths per 100,000 in absolute value at the four placebo
event times, with standard errors near $2$, against a largest placebo coefficient of $2.9$. With only the female and
Hispanic shares dated in 2013 they are between $0.2$ and $0.3$. The test and the shock terms disagree for the reason given after Proposition~\ref{prop:timetest}. The female share
records mortality with a very small weight. One more death per 100,000 adults among men raises it by about half a
thousandth of a percentage point, so the weight $\chi$ that sets the size of the shock term is tiny. The share is
measured almost without error, so the noise $\Var(\zeta)$ is tiny as well, and the test detects even this weight in a
sample of 2,200 counties. This is the case of Proposition~\ref{prop:timetest}(iv), a rejection whose shift is a small
fraction of the benchmark, and it is the reason to report the shock terms with the test.

\begin{table}[!htb]
\centering
\small
\begin{tabular}{lccc}
\hline
\multicolumn{4}{l}{\emph{Test of Proposition~\ref{prop:timetest}: 2013 value given the 2009 values and treatment, four windows}}\\
covariate & Wald & df & $p$ \\
\hline
female share & $14.5$ & 4 & $0.006$ \\
white share & $7.0$ & 4 & $0.138$ \\
Hispanic share & $10.8$ & 4 & $0.028$ \\
unemployment rate & $1.5$ & 4 & $0.827$ \\
poverty rate & $1.0$ & 4 & $0.908$ \\
median income & $6.1$ & 4 & $0.195$ \\
all six & $48.0$ & 24 & $0.003$ \\
\hline
\multicolumn{4}{l}{\emph{Event study with each dating of the covariates}}\\
dating & largest $|$placebo$|$ & average effect, $e=0,\dots,5$ & set at $M=0.5$ \\
\hline
no covariates & $4.13$ & $-0.70$ $(2.02)$ & $[-15.1,\,12.7]$ \\
all 2013 \citep{Baker2025} & $2.88$ & $0.46$ $(2.36)$ & $[-13.8,\,15.7]$ \\
all 2009 & $2.88$ & $4.49$ $(2.10)$ & $[-9.8,\,19.1]$ \\
unemployment 2009, others 2013 & $2.91$ & $3.72$ $(2.12)$ & \\
female and Hispanic 2009, others 2013 & $3.17$ & $0.75$ $(2.35)$ & \\
\hline
\end{tabular}
\caption{Medicaid expansion and county adult mortality (deaths per 100,000 adults aged 20--64), 2014 expansion counties
against counties not expanded by 2019. Upper panel, Wald tests with county-clustered standard errors. Lower panel, outcome
regression with base year 2013 and 2013 population weights, standard errors in parentheses, and the relative-magnitude set
of \citet{RR} for the average effect at $M=0.5$.}
\label{tab:medicaid}
\end{table}

Table~\ref{tab:medicaid}, lower panel, reports the average effect over event times
$0$ to $5$ and the largest placebo coefficient under several datings. Dating every covariate in 2009 raises the average
effect from $0.46$ to $4.49$ deaths per 100,000. For scale, \citet{Baker2025} report an effect of about $-2.5$ in the
first year in their 2$\times$2 design, a different estimand from our six-year average. Meanwhile the largest placebo coefficient is $2.88$ either way, so the pre-trend check does not
reveal the change. Changing the dating of one covariate at a time locates it in the unemployment rate. Dating it alone in
2009 gives $3.72$, and dating any other covariate alone in 2009 moves the estimate by less than $0.4$. The unemployment
rate passes the test most clearly, and its 2009 value is the peak of the recession, so the 2009 dating replaces a
covariate that is valid for the benchmark by a stale measurement. In terms of the theory, the characteristic that
belongs in the role of $X_0$ is the county's persistent labor-market condition. The 2013 rate measures it without
recording the outcome, as the test shows, and the 2009 rate measures it with a large cyclical error that has nothing to
do with mortality. Both values are then functions of $X_0$ and of noise unrelated to the outcome, the second case of
Corollary~\ref{cor:contrast}, and switching between them moves the trend terms, and with them the estimate, while the
placebo coefficients stay where they were. Dating early protects the benchmark only when the early value measures the
same characteristic about as well as the late one does. The dating that the test
suggests keeps the 2013 values of the four covariates that pass and uses the 2009 values of the female and Hispanic
shares. It gives an average effect of $0.75$ (s.e.\ $2.35$), close to the $0.46$ of the 2013 dating, so the conclusion of
\citet{Baker2025} that the mortality effect is small and imprecise survives. The relative-magnitude sets at $M=0.5$
of \citet{RR} contain zero under every dating.

\subsection{Telephone automation, staggered adoption}
\label{sec:phone}

The third application is taken from the survey. \citet{FeigenbaumGross2024} estimate the effect of the mechanization of
local telephone exchanges, the cutover from manual to dial service, on the employment of young women as telephone
operators in U.S.\ cities. We rebuild their city panel from the replication package, follow their
sample restrictions, and reproduce their Figure~3(A) to two decimals. The panel has 2,845 cities observed in the
censuses of 1910 to 1940, and 307 of them cut over between 1917 and 1940. The outcome is the log of one plus the number
of young, white, U.S.-born women working as telephone operators in the telephone industry. The event study has city and
year effects and bins event time in decades, with the decade before the cutover omitted, so the two placebo coefficients
compare the outcome 10 to 20 and more than 20 years before the cutover with the decade before it. Standard errors are
clustered by city.

The covariates of \citet{FeigenbaumGross2024} are the log number of employed young women and the log population of the
city, the latter interacted with year, both measured in the census year of each observation. This is dating (P4). Each
pre-cutover census enters as the end of a placebo window, so the covariate of that census records the same growth of
the city that the outcome records. Both covariates are also observed in 1910, the first census of the panel, and we
refit the event study with the 1910 values interacted with year. A linear version of the test of Proposition~\ref{prop:timetest},
as in the Medicaid application, rejects decisively. We regress each decade change in the outcome that ends before the cutover, 1910--1920, 1920--1930 and
1930--1940, on cutover-decade indicators, the 1910 covariates and their squares, and the two covariates of the closing
census, all specific to the window. The census-year employment of young women predicts the change in every window, with
$t$-statistics between $3.1$ and $8.5$, and the joint Wald statistic for the six coefficients on the census-year covariates
is $937$ on six degrees of freedom.

\begin{table}[!htb]
\centering
\small
\begin{tabular}{lccccc}
\hline
 & \multicolumn{2}{c}{placebo coefficient} & & & \\
dating & $<-20$ & $-20$ to $-10$ & average effect & breakdown $M$ & set at $M=0.5$ \\
\hline
no covariates & $-0.31$ & $-0.21$ & $-0.58$ $(0.07)$ & $0.95$ & $[-0.93,\,-0.24]$ \\
census year (published) & $-0.13$ & $-0.06$ & $-0.62$ $(0.06)$ & $2.15$ & $[-0.82,\,-0.41]$ \\
1910 & $-0.31$ & $-0.20$ & $-0.45$ $(0.08)$ & $0.70$ & $[-0.80,\,-0.10]$ \\
\hline
shock term, census year & $0.17$ $(0.03)$ & $0.14$ $(0.02)$ & & & \\
\hline
\end{tabular}
\caption{Telephone automation and the log number of young women employed as telephone operators, 2,845 U.S.\ cities,
censuses of 1910--1940. Event-time bins in decades relative to the cutover, the decade before it omitted, standard errors
clustered by city in parentheses. The average effect is the mean of the three post-cutover bins. The breakdown value is
the largest $M$, on a grid of $0.05$, at which the relative-magnitude set of \citet{RR} for the average effect
excludes zero. Shock terms of Proposition~\ref{prop:shock}, linear version, with city-cluster bootstrap standard errors
from 199 draws.}
\label{tab:phone}
\end{table}

Table~\ref{tab:phone} reports the event study under each dating. Without covariates the placebo coefficients are $-0.31$
and $-0.21$, so operator employment was rising in the cities that were about to cut over. The census-year covariates
shrink them to $-0.13$ and $-0.06$, while the 1910 covariates leave them at $-0.31$ and $-0.20$. We estimate the shock terms as in Proposition~\ref{prop:shock}, with the regression in place of the contrast. The
outcome is first residualized on the event-time indicators, the city effects and the year-specific effects of the 1910
covariates, and the published event study, with the census-year covariates, is then fitted to the residual. Its placebo
coefficients are the shock terms. They are $0.17$ and $0.14$, and they equal the gaps between the two sets of placebo
coefficients, $0.18$ and $0.14$. On the scale of the trend they are $-0.17$ and $-0.14$, so they lower the rising
pre-trend toward zero, the direction of Proposition~\ref{prop:latesign}. Almost the whole shift comes from city
population interacted with year. With census-year population as the only covariate the placebo coefficients are already
$-0.13$ and $-0.06$. Growing cities
hired more operators, and the census-year population absorbs that growth in the placebo windows. The late dating
therefore masks a pre-trend that the early dating reveals, and unlike the setting of
Corollary~\ref{cor:contrast} it also moves the estimate. The average post-cutover effect is $-0.62$ with the census-year covariates and $-0.45$ with the
1910 covariates. The two effects combine in the honest band. The breakdown value of the relative-magnitude bound is
$2.15$ under the published dating and $0.70$ under the early one, so the late covariates make the result look three
times as robust to violations of parallel trends as the early covariates do. The conclusion that automation reduced
operator employment survives, since the set at $M=0.5$ excludes zero under every dating, but the size of the
effect and its robustness are both overstated by the late dating.

The three applications answer different questions. The China design is the most complete test of the theory. The same
covariates are observed at three dates, every placebo slope falls as the dating moves later, the signs of the shock
terms agree with Proposition~\ref{prop:latesign}, and the bound of Proposition~\ref{prop:sens} recovers the early
benchmark from the late data alone. The automation design applies the tools to a published study. It is
staggered, it uses the dating that the survey finds most often, and its late covariates raise the breakdown value from
$0.70$ to $2.15$, although the conclusion that automation reduced operator employment holds under either dating. The
Medicaid design shows why the test is paired with the shock terms. Two covariates fail the test, their estimated shock
terms show that the failure barely moves the benchmark, and the dating that moves the estimate is of a covariate that
passes. The rule is
therefore to date early the covariates whose shock terms are large, not every covariate, and to report the estimate and
the breakdown value under each dating.

\section{Conclusion}
\label{sec:conclusion}

Benchmarking post-treatment violations of conditional parallel trends by conditional pre-trends requires covariates that
predate the pre-period. A covariate measured at the end of the pre-period partly records the changes in the outcome during the pre-period,
and conditioning on it adds to every placebo estimate a shock term that the post-period trend does not share. The shock
term is separate from any bias in the estimate, and it can distort the benchmark when the estimate is untouched. In the
usual case the placebo estimates are pulled below the post-period trend, so a positive trend is masked and the honest
band is too short. The problem arises with binary, continuous and staggered treatments, and in staggered designs it
follows from the common practice of fixing covariates at each cohort's base period.

When a covariate is observed at an early and a late date, the condition can be tested and the shock terms estimated, both
by partially linear regressions that are cheap to run alongside any covariate-adjusted event study. When only the late
date exists, a single correlation bounds the shock terms. In the China-shock design the late dating hides the most recent
pre-trend, the estimated shock terms account for the change in the benchmark, and a modest value of the sensitivity
parameter recovers the predetermined benchmark. In the Medicaid design the test rejects for covariates whose shock terms
are small, and the dating that moves the estimate is of a covariate that passes. In the automation design the
published dating hides most of the pre-trend and raises the breakdown value from $0.70$ to $2.15$, although the
conclusion that automation reduced operator employment holds under either dating. Late dating is common.
Of 90 recent published difference-in-differences studies that adjust their main specification for time-varying covariates, 77 date at least
one covariate after some placebo window, and 71 of these record an earlier value with which the condition could be
tested. How much the dating matters differs across studies, and the test and the shock terms tell a researcher which case
applies.

\clearpage
\appendix

\section{Notation and a step-by-step procedure}
\label{app:steps}

\begin{table}[htbp]
\centering
\small
\begin{tabular}{p{0.24\textwidth}p{0.56\textwidth}l}
\hline
symbol & meaning & first used \\
\hline
$D$, $X_0$, $\tilde X$ & treatment, predetermined covariate, covariate used for adjustment & \S\ref{sec:setup} \\
$s=1,\dots,K$, post & pre-period windows and the post-period window & \S\ref{sec:setup} \\
$\beta(W;\tilde X)$, $\beta^\omega$, $\beta_d$ & within-$\tilde X$ contrast, regression summary, slope at dose $d$ & \S\ref{sec:setup} \\
$\beta_s(\tilde X)$, $\beta_{\mathrm{post}}(\tilde X)$ & adjusted placebo estimand of window $s$, post-period within-cell trend & \S\ref{sec:setup} \\
$m_{\tilde X}$ & adjusted post-period contrast, $\ATT+\beta_{\mathrm{post}}(\tilde X)$ & \S\ref{sec:setup} \\
$M$, $M^*$, $M^\circ_{\tilde X}$ & trend budget, breakdown value, budget needed for coverage & \S\ref{sec:setup}, \eqref{eq:MX} \\
$Z$, $V$, $\zeta$ & late covariate, cumulative pre-period change, noise in $Z$ & Ass.~\ref{ass:lategen} \\
$\tau_s$, $u_s$, $U$ & within-cell trend, shock of window $s$, sum of pre-period shocks & Ass.~\ref{ass:lategen} \\
$T_s$, $\varepsilon_s$ & trend term and shock term of window $s$ & Prop.~\ref{prop:lategen} \\
$A$, $\Psi$ & scalar index of $(D,X_0)$ and the function that maps $(A,U)$ to $Z$ & Prop.~\ref{prop:latesign} \\
$b$, $\delta$, $\chi$, $\kappa$ & effect of $X_0$ and of $D$ on the trend, weight of $V$ in $Z$, $1+\chi Kb$ & Ass.~\ref{ass:linear} \\
$\gamma$, $\Lambda$, $\Xi_s$ & slope of $D$ on $Z$, residual variance of $D$, $\Cov(u_s,U)$ & Prop.~\ref{prop:late} \\
$G$, $g$, $X_t$ & first treated period, cohort, covariate in period $t$ & \S\ref{sec:stagger} \\
$\vartheta_s$ & partially linear coefficient of $Z$ in the timing test & Prop.~\ref{prop:timetest} \\
$R_s$ & residual pre-period change given $(D,X_0)$ & Prop.~\ref{prop:shock} \\
$\rho_s$, $\bar\rho$, $B_s$ & sensitivity correlation, its bound, identified scale & Prop.~\ref{prop:sens} \\
\hline
\end{tabular}
\caption{Notation of the main text.}
\label{tab:notation}
\end{table}

Table~\ref{tab:notation} collects the notation. The rest of this appendix lists the steps that take a researcher from
the data to the reported bands, in the order in which they are carried out. The data are units $i=1,\dots,n$, possibly in
clusters and with weights, a treatment $D_i$, covariates, the change after treatment and the placebo changes
$\Delta Y_{is}$ for windows $s=1,\dots,K$.

\begin{enumerate}[label=\textbf{\arabic*.},leftmargin=*,itemsep=3pt]

\item \textbf{Date every covariate.} For each covariate record the date at which it is measured in the main
specification and classify the dating as (P1) to (P4) of Section~\ref{sec:stagger}. A covariate measured before the
start of the earliest placebo window needs no further work. For each of the others record whether the data contain an
earlier value of the same variable, measured before that window.

\item \textbf{Build the placebo changes.} For each window that ends before treatment form the change in the outcome and
attach to it the treatment, covariates and weights of the post-period comparison, so that every placebo comparison
shares their distribution.

\item \textbf{Estimate the placebo and post-period slopes.} Compute the adjusted slopes with the dating of the main
specification. The partially linear slope of \citet{Robinson1988} is the residual of the outcome change regressed on the
residual of the treatment, both residualized on the covariates. The residuals can come from least squares on a series in
the covariates, such as their levels and squares, or from cross-fitted flexible regressions \citep{DML2018} with folds
formed by cluster. In the China application they are lasso fits on splines and pairwise interactions, with the median
over $25$ splits.

\item \textbf{Test the timing of each late covariate.} Where an early value $X_0$ exists, residualize the late value $Z$
and each placebo change on $(D,X_0)$, including the square of the treatment when it is continuous. The coefficient
$\hat\vartheta_s$ is the sum of the products of the two residuals divided by the sum of the squared residuals of $Z$.
Stack the influence functions of the $\hat\vartheta_s$ across windows and covariates, sum them within clusters, and form
the Wald statistic with the generalized inverse of their covariance and degrees of freedom equal to its rank. In a
regression event study the linear version adds the late covariates to an event study that already controls for the
early ones interacted with period, and tests their coefficients.

\item \textbf{Estimate the shock terms.} For each placebo window form the residual $\hat R_s$ of the change on
$(D,X_0)$, and compute its adjusted slope on the treatment given $Z$, exactly as the placebo slope of step 3 with
$\hat R_s$ in place of the change. In a regression event study, fit the event study with the early covariates, take its
residual, and refit the published event study to the residual. Standard errors can treat the first-step residuals as
known or come from a cluster bootstrap of both steps.

\item \textbf{Judge the size of the shock terms.} Compare each shock term with the standard error of the largest placebo
estimate (Section~\ref{sec:test}). Date early the covariates whose shock terms are of that order or more, and keep the
late values of the others.

\item \textbf{Bound the shock terms when no early value exists.} Compute $B_s$ as the standard deviation of the residual
of the placebo change given the treatment alone divided by the standard deviation of the residual of the treatment given
$Z$. For a grid of $\bar\rho$ replace each placebo slope by the larger of $|\hat\beta^\omega_s(Z)|$ and
$|\hat\beta^\omega_s(Z)+\bar\rho B_s|$, recompute the band, and report the value of $\bar\rho$ at which its conclusion
changes. Choose the values of $\bar\rho$ to emphasize as in Section~\ref{sec:test}.

\item \textbf{Report.} For each dating used, report the estimate, the largest placebo estimate, the band at the chosen
$M$ and the breakdown value $M^*$, together with the test, the shock terms or the sensitivity bound. A plot of the
placebo slopes under each dating, as in Figure~\ref{fig:china}, shows at a glance whether the dating moves them.

\end{enumerate}

\section{Proofs}
\label{sec:proofs}

Throughout, $\beta$ denotes any of the contrasts of Section~\ref{sec:setup} and Remark~\ref{rem:continuous}. Each is linear
in its first argument and depends on that argument only through its conditional mean given $(D,\tilde X)$.

\begin{proof}[Proof of Proposition~\ref{prop:lategen}]
(i) Since $u_s$ is independent of $(D,X_0)$, $\E[\Delta Y_s(0)\mid D,X_0]=\alpha_s+\tau_s(D,X_0)$, and the constant drops out
of every contrast. (ii) By linearity, $\beta(\Delta Y_s(0);Z)=\beta(\alpha_s;Z)+\beta(\tau_s(D,X_0);Z)+\beta(u_s;Z)$, and
the first term is zero. (iii) The first claim follows from (ii). If $Z=\psi(X_0,\zeta)$, then $\E[u_s\mid D,Z]=\E[\E[u_s\mid
D,X_0,\zeta]\mid D,Z]=0$, because $u_s$ is independent of $(D,X_0)$ and $\zeta$ of $(D,X_0,u_s)$, so $\varepsilon_s=0$.
Joint independence of the shocks from $(D,X_0)$ and independence of $u_{\mathrm{post}}$ from $(u_1,\dots,u_K)$ make
$u_{\mathrm{post}}$ independent of $(D,X_0,u_1,\dots,u_K)$, hence of $(D,X_0,u_1,\dots,u_K,\zeta)$. Since $Z$ is a
function of these, $\E[u_{\mathrm{post}}\mid D,Z]=0$ and $\varepsilon_{\mathrm{post}}=0$. \end{proof}

\begin{proof}[Proof of Corollary~\ref{cor:contrast}]
For a function $h_s(D)$ of a binary treatment, $\E[h_s(D)\mid D=1,\tilde X]-\E[h_s(D)\mid D=0,\tilde X]=h_s(1)-h_s(0)$ for
every $\tilde X$, so $\beta(h_s(D);Z)=\beta(h_s(D);X_0)$, and $\Cov(D,h_s(D)\mid\tilde X)=(h_s(1)-h_s(0))\Var(D\mid\tilde X)$,
so $\beta^\omega(h_s(D);\tilde X)=h_s(1)-h_s(0)$ as well. Hence $T_s=\beta_s(X_0)$ for every window and both contrasts,
and $\beta_s(Z)=\beta_s(X_0)+\varepsilon_s$ by Proposition~\ref{prop:lategen}(ii). With $\varepsilon_{\mathrm{post}}=0$ from
Proposition~\ref{prop:lategen}(iii), $\beta_{\mathrm{post}}(Z)=\beta_{\mathrm{post}}(X_0)$, and since
$m_{\tilde X}=\ATT+\beta_{\mathrm{post}}(\tilde X)$ for every $\tilde X$, $m_Z=m_{X_0}$. The last claim is
Proposition~\ref{prop:lategen}(iii).
\end{proof}

\begin{proof}[Proof of Proposition~\ref{prop:latesign}]
(i) Let $k_s(U)\coloneqq\E[u_s-u_{\mathrm{post}}\mid U]$. The shocks are jointly independent of $(D,X_0)$, so
$\E[u_s-u_{\mathrm{post}}\mid D,X_0,U]=k_s(U)$, and since $Z$ is a function of $(D,X_0,U)$, $\Cov(D,u_s-u_{\mathrm{post}}\mid Z)=\Cov(D,k_s(U)\mid Z)$. Given $Z=z$,
$U=\Psi^{-1}(A;z)$ is a decreasing function of $A$, so $k_s(U)$ is a nonincreasing function of $A$. Because $(D,A)$ is
independent of $U$, $\E[D\mid A,Z]=\E[D\mid A,U]=\E[D\mid A]$, which is nondecreasing in $A$. Hence $\Cov(D,u_s-u_{\mathrm{post}}\mid
Z)=\Cov(\E[D\mid A],k_s(\Psi^{-1}(A;Z))\mid Z)\le0$ by Chebyshev's association inequality, and averaging over $Z$ and dividing
by $\E[\Var(D\mid Z)]$ gives $\varepsilon^\omega_s\le\varepsilon^\omega_{\mathrm{post}}$. The decomposition and the
stationary case follow from Proposition~\ref{prop:lategen}(iii) applied to $\beta^\omega$. (ii) Write $\beta_s$ and
$\beta_{\mathrm{post}}$ for the two slopes. If $\beta_{\mathrm{post}}>0$ and $-\beta_{\mathrm{post}}<\beta_s\le
\beta_{\mathrm{post}}$ for every $s$, then $\max_s|\beta_s|\le\beta_{\mathrm{post}}$, with strict inequality when every
$\beta_s<\beta_{\mathrm{post}}$. If $\beta_{\mathrm{post}}<0$, then $\beta_s\le\beta_{\mathrm{post}}<0$ and
$|\beta_s|\ge|\beta_{\mathrm{post}}|$. (iii) The same argument with $\E[D\mid A]$ nonincreasing gives $\Cov(D,u_s-u_{\mathrm{post}}\mid
Z)\ge0$. \end{proof}

\begin{proof}[Proof of Proposition~\ref{prop:late}]
We prove a version that allows an arbitrary covariance of the shocks across windows. Write $U\coloneqq\sum_{s=1}^Ku_s$ and
$\Xi_s\coloneqq\Cov(u_s,U)$ for $s\in\{1,\dots,K,\mathrm{post}\}$, and keep the other parts of Assumption~\ref{ass:linear}. We show
\[
  \beta_s(X_0)=\delta ,\qquad
  \beta_s(Z)=\frac{\delta }{\kappa}-\frac{\chi\gamma}{\Lambda}\Bigl(\Xi_s-\frac{b\chi}{\kappa}\Var(U)\Bigr),\qquad
  \beta_{\mathrm{post}}(Z)-\beta_s(Z)=\frac{\chi\gamma\,(\Xi_s-\Xi_{\mathrm{post}})}{\Lambda}.
\]
With $\tilde X=X_0$ the shock $u_s$ is uncorrelated with the regressors, so the projection coefficient on $D$ is $\delta $. Write
$V=\bar\alpha+KbX_0+K\delta D+U$ with $\bar\alpha=\sum_{s=1}^K\alpha_s$. Then $Z=\chi\bar\alpha+\kappa X_0+\chi K\delta D+\chi U$, so
$X_0=(Z-\chi\bar\alpha-\chi K\delta D-\chi U)/\kappa$ and
\[
  \Delta Y_s(0)=\text{const}+\frac{b}{\kappa}Z+\frac{\delta }{\kappa}D+e_s,\qquad e_s\coloneqq u_s-\frac{b\chi}{\kappa}U,
\]
using $\delta -b\chi K\delta/\kappa=\delta /\kappa$. The coefficient on $D$ is therefore $\delta /\kappa$ plus the coefficient on $D$ in the
projection of $e_s$ on $(1,D,Z)$. By the Frisch--Waugh theorem that coefficient is $\Cov(e_s,\tilde D)/\Lambda$, with
$\tilde D=D-\E D-\gamma(Z-\E Z)$ and $\Var(\tilde D)=\Lambda$. Here $\Lambda>0$ because $D$ is a linear function of $Z$ only if it is a
linear function of $X_0$. Since $e_s$ is uncorrelated with $D$ and $X_0$,
$\Cov(e_s,\tilde D)=-\gamma\Cov(e_s,Z)=-\gamma\chi\Cov(e_s,U)=-\gamma\chi\bigl(\Xi_s-b\chi\Var(U)/\kappa\bigr)$, which gives the
middle display. The same computation for $u_s$ alone gives the shock term $\varepsilon_s=-\chi\gamma\Xi_s/\Lambda$. The term in
$\Var(U)$ is common to all windows and cancels in the difference. Under Assumption~\ref{ass:linear}, $\Xi_s=\sigma_u^2$ for
$s\le K$, $\Xi_{\mathrm{post}}=0$ and $\Var(U)=K\sigma_u^2$, and \eqref{eq:lateiid} follows using $\sigma_u^2-b\chi
K\sigma_u^2/\kappa=\sigma_u^2/\kappa$. The expression for $M^\circ_Z$ follows from \eqref{eq:MX} after multiplying through by
$\kappa$, and its denominator vanishes as $\delta \to\chi\gamma\sigma_u^2/\Lambda$ while the numerator tends to
$\kappa\chi\gamma\sigma_u^2/\Lambda\ne0$. With $b=0$, $\kappa=1$ and the second expression in \eqref{eq:lateiid} equals $\delta $.
\end{proof}

\begin{proof}[Proof of Proposition~\ref{prop:stagger}]
Each comparison of cohort $g$ with the never-treated units is a two-group design to which Assumption~\ref{ass:lategen}
applies. (i) The covariate $X_1$ plays the role of $X_0$, and Proposition~\ref{prop:lategen}(i) applies. (ii) Under (P2)
every comparison of cohort $g$ conditions on $X_{g-1}=\psi_{g-1}(X_1,Y_{g-1}(0)-Y_1(0),\zeta_{g-1})$, a late covariate in
the sense of Assumption~\ref{ass:lategen} with $V$ the sum of the cohort's pre-period changes, so
Propositions~\ref{prop:lategen} and~\ref{prop:latesign} apply. An aggregate at event time $e$ is
$\sum_gw_g\beta_{g,g+e}$ with $w_g\ge0$, and with the same weights at a placebo event time and at the post-period event
time the inequality of Proposition~\ref{prop:latesign}(i) is preserved term by term. (iii) Under (P3) the comparison over
the window ending at $t$ conditions on $X_{t-1}$, a function of $(X_1,u_2,\dots,u_{t-1},\zeta_{t-1})$ and of $D$ through
the trends. With shocks independent across periods, $u_t$ is independent of these, so $\E[u_t\mid D,X_{t-1}]=0$ and the
shock term is zero, while the trend term is $\beta(\tau_t(D,X_1);X_{t-1})$. \end{proof}

\begin{proof}[Proof of Proposition~\ref{prop:timetest}]
(i) If $Z=\psi(X_0,\zeta)$, then $\E[u_s\mid D,X_0,Z]=0$, because $u_s$ is independent of $(D,X_0)$ and $\zeta$ is
independent of $(D,X_0,u_s)$, and $\E[\Delta Y_s\mid D,X_0,Z]=\alpha_s+\tau_s(D,X_0)=\E[\Delta Y_s\mid D,X_0]$. (ii) Under $H_0$,
$\E[u_s\mid D,X_0,Z]=0$, so $\E[u_s\mid D,Z]=0$ and $\varepsilon_s=0$. With $u_{\mathrm{post}}$ independent of
$(u_1,\dots,u_K)$, $\varepsilon_{\mathrm{post}}=0$ by Proposition~\ref{prop:lategen}(iii), and with stationary trends
$\beta_s(Z)=\beta_{\mathrm{post}}(Z)=T$, so $M^\circ_Z=1$ by \eqref{eq:MX}. (iii) Under $H_0$, $\E[(Z-\E[Z\mid D,X_0])\Delta
Y_s]=\E[(Z-\E[Z\mid D,X_0])\,\E[\Delta Y_s\mid D,X_0]]=0$. In the linear design the residual of $Z$ given $(D,X_0)$ is $\chi
U+\zeta$ and that of $\Delta Y_s$ is $u_s$, which gives the formula. (iv) The first claim is the contrapositive of (ii).
In the linear design the shock term is $\varepsilon_s=-\chi\gamma\Xi_s/\Lambda$, as shown in the proof of
Proposition~\ref{prop:late}, and the argument there is unchanged by adding $\zeta$ to $Z$. It is zero when
$\gamma=\Cov(D,Z)/\Var(Z)=0$, while $\vartheta_s\ne0$ by (iii) when $\chi\Xi_s\ne0$. \end{proof}

\begin{proof}[Proof of Proposition~\ref{prop:shock}]
Since the shocks are independent of $(D,X_0)$, $\E[\Delta Y_s\mid D,X_0]=\alpha_s+\tau_s(D,X_0)$ and $R_s=u_s$. Then
$\varepsilon_s=\beta(u_s;Z)=\beta(R_s;Z)$, and $T_s=\beta_s(Z)-\varepsilon_s$ by Proposition~\ref{prop:lategen}(ii). \end{proof}

\begin{proof}[Proof of Proposition~\ref{prop:sens}]
By definition $\E[\Cov(D,u_s\mid Z)]=\E[\tilde Du_s]$ and $\E[\Var(D\mid Z)]=\E[\tilde D^2]$. Since $\E[\tilde D\,h(Z)]=0$ for
every square-integrable $h$, $\E[\tilde Du_s]=\E[\tilde D\tilde u_s]$ with $\tilde u_s\coloneqq u_s-\E[u_s\mid Z]$, and $\E\tilde
D=\E\tilde u_s=0$, so $\varepsilon^\omega_s=\Cov(\tilde D,\tilde u_s)/\Var(\tilde D)=\rho_s\,\mathrm{sd}(\tilde u_s)/\mathrm{sd}(\tilde D)$.
For the bound, $\Var(\tilde u_s)\le\Var(u_s)$. Moreover $\Delta Y_s-\E[\Delta Y_s\mid D]=(\tau_s(D,X_0)-\E[\tau_s(D,X_0)\mid
D])+u_s$, because $\E[u_s\mid D]=0$, and the two terms are uncorrelated because $u_s$ is independent of $(D,X_0)$. Hence
$\Var(u_s)\le\Var(\Delta Y_s-\E[\Delta Y_s\mid D])$ and $|\varepsilon^\omega_s|\le|\rho_s|B_s$. \end{proof}

\end{document}